\documentclass[aps,prab,twocolumn,superscriptaddress,floatfix]{revtex4-1}

\usepackage{amssymb}
\usepackage{lipsum}
\usepackage{amssymb}
\usepackage{bbold}
\usepackage{amssymb}
\usepackage{graphicx}
\usepackage{graphics}
\usepackage{amsmath}
\usepackage{siunitx}
\usepackage{placeins}
\usepackage{hyperref}
\usepackage[dvipsnames]{xcolor}
\usepackage{wasysym}
\usepackage{soul}
\usepackage{float}
\usepackage{array}
\usepackage{tabulary}
\usepackage{caption}
\usepackage{fancyhdr}
\usepackage{wrapfig}

\usepackage[normalem]{ulem}
\usepackage{graphicx}
\usepackage{color,soul}
\usepackage{subfigure}
\usepackage{dcolumn}
\usepackage{bm}
\usepackage{hyperref}
\usepackage{url}
\usepackage{multirow}
\hypersetup{
    colorlinks=true,
    linkcolor=blue,
    filecolor=blue,      
    urlcolor=blue,
    citecolor=blue,
}

\usepackage[makeroom]{cancel}

\begin{document}

\title{Thermal and electric field driven breakdown precursor formation on metal surfaces} 

    \author{\firstname{Ryo} \surname{Shinohara}}
    \affiliation{Department of Electrical and Computer Engineering, Michigan State University, MI 48824, USA}
    \affiliation{Department of Physics and Astronomy, Michigan State University, East Lansing, MI 48824, USA}
    \affiliation{Theoretical Division, Los Alamos National Laboratory, Los Alamos, NM 87545, USA}
    \author{\firstname{Soumendu} \surname{Bagchi}}
    \affiliation{Theoretical Division, Los Alamos National Laboratory, Los Alamos, NM 87545, USA}
    \author{\firstname{Evgenya} \surname{Simakov}}
    \affiliation{Accelerator Operations and Technology Division, Los Alamos National Laboratory, Los Alamos, NM 87545, USA}
    \author{\firstname{Sergey V.} \surname{Baryshev}}
	\affiliation{Department of Electrical and Computer Engineering, Michigan State University, MI 48824, USA}
	\affiliation{Department of Chemical Engineering and Material Science, Michigan State University, MI 48824, USA}
    \author{\firstname{Danny} \surname{Perez}}
    \affiliation{Theoretical Division, Los Alamos National Laboratory, Los Alamos, NM 87545, USA}
    
\begin{abstract}
    The phenomenon of electric breakdown poses serious challenge to the design of devices that operate under high gradient environments. Experimental evidence often points towards breakdown events that are accompanied by elevated temperatures and dark current spikes, presumably due to high-asperity nano-structure formation that enhances the local electric field and triggers a runaway process. However, the exact mechanistic origin of such nano-structures under typical macroscopic operational conditions of electric gradient and magnetic-field-mediated heating remains poorly understood. In this work, a model is presented that describes the evolution of a typical copper surface under the combined action of the electric fields and elevated temperatures. Using a mesoscale curvature-driven growth model, we show how the copper surface can undergo a type of dynamical instability that naturally leads to the formation of sharp asperities in realistic experimental conditions.
    Exploring the combined effect of fields and temperature rise, we identify critical regimes that allow for the formation of breakdown precursors. These regimes strongly resonate with previous experimental findings on breakdown of copper electrodes, hence suggesting surface diffusion to be a crucial breakdown precursor mechanism.   
\end{abstract}

\maketitle

\section{Introduction}\label{S:intro}
Application of high electric field are commonplace in modern vacuum technologies ranging from space propulsion to powerful linear accelerators \cite{14, simakov2018advances}. In this latter use case, next-generation accelerator technologies aim to operate in large field gradient regimes, over 100 MV/m \cite{15,16}. However, such high field are known to lead to vacuum arcs 
launched by so-called breakdown (BD) events. These events not only perturb normal operation, but also cause damages to the devices, ultimately limiting the maximum achievable fields. In spite of the fact that BD has been a known phenomenon for over a century, the detailed mechanism behind BD nucleation and evolution remain poorly understood, in great part because BD events destroys their precursors.

Previous experimental, theoretical, and computational efforts \cite{10,11,12,5,6} have proposed essential processes, later categorizing them into two major groups: 1) processes that lead to the formation of (i.e. nucleate) BD events and 2) processes that take place during the BD event, including its terminal stage. It is hypothesized that a BD event likely nucleates through the formation of geometrical asperities. These asperities can locally enhance the applied gradient, triggering a runaway process eventually leading to material evaporation and the terminal arc formation. However, the exact mechanisms behind the formation of such nano-structures on mirror-quality polished surface with nanoscale roughness remain unknown.

What is known is that critical breakdown rates strongly depend on applied electric and microstructural variations \cite{simakov2018advances}. While plastic deformation driven by electric fields \cite{Hebrew2019, pohjonen2011dislocation, atomistic-coupling2022} have been computationally explored as candidate mechanisms to understand BD phenomenon, the applied fields considered in  these studies (e.g., 5-10 GV/m) have been mostly orders of magnitude higher than those applied in experiments (e.g. 100-200 MV/m). On the other hand, the breakdown rate is known to also be very sensitive to the temperature \cite{cryo-breakdown, dobert, grudiev}. While reducing temperature immediately decreases the BDR \cite{cryo-breakdown}, increasing the operating temperature above room temperature by only 50 K quickly increases rate of breakdown and essentially limits the practical gradients \cite{cryo-breakdown, dobert}. When not actively cooled, the operating temperature naturally grows due to a process called pulse heating, where it is capable of producing temperature rise in the order of several hundred degrees \cite{RFPulsedHeating}.
These results from the effects associated with the magnetic field component of the radio-frequency (rf) signal and the duty cycle (rf pulse length vs. repetition rate). The combined pulsed heating and peak electric fields set an upper limits on the gradients that can be used in normal operation
if a sufficiently low breakdown rate is to be maintained.

The synergy between the electric gradient and temperature, as well as the effect of nominal surface geometries are relatively unexplored in previous analyses \cite{Hebrew2019}. Here, we explore surface diffusion as a possible mechanisms enabling BD nucleation under nominal experimental $E$-field and heating conditions. By considering the competition between surface tension and the external electric and thermo-mechanical driving forces, we show that moderate initial surface perturbations are capable of triggering a structural instability through surface diffusion, leading to the formation of BD precursors at electric field and temperature gradients comparable to experimental measurements reported in operating accelerators.

\section{Problem formulation}\label{Fab}
\begin{figure}
	\includegraphics[width=8.6cm]{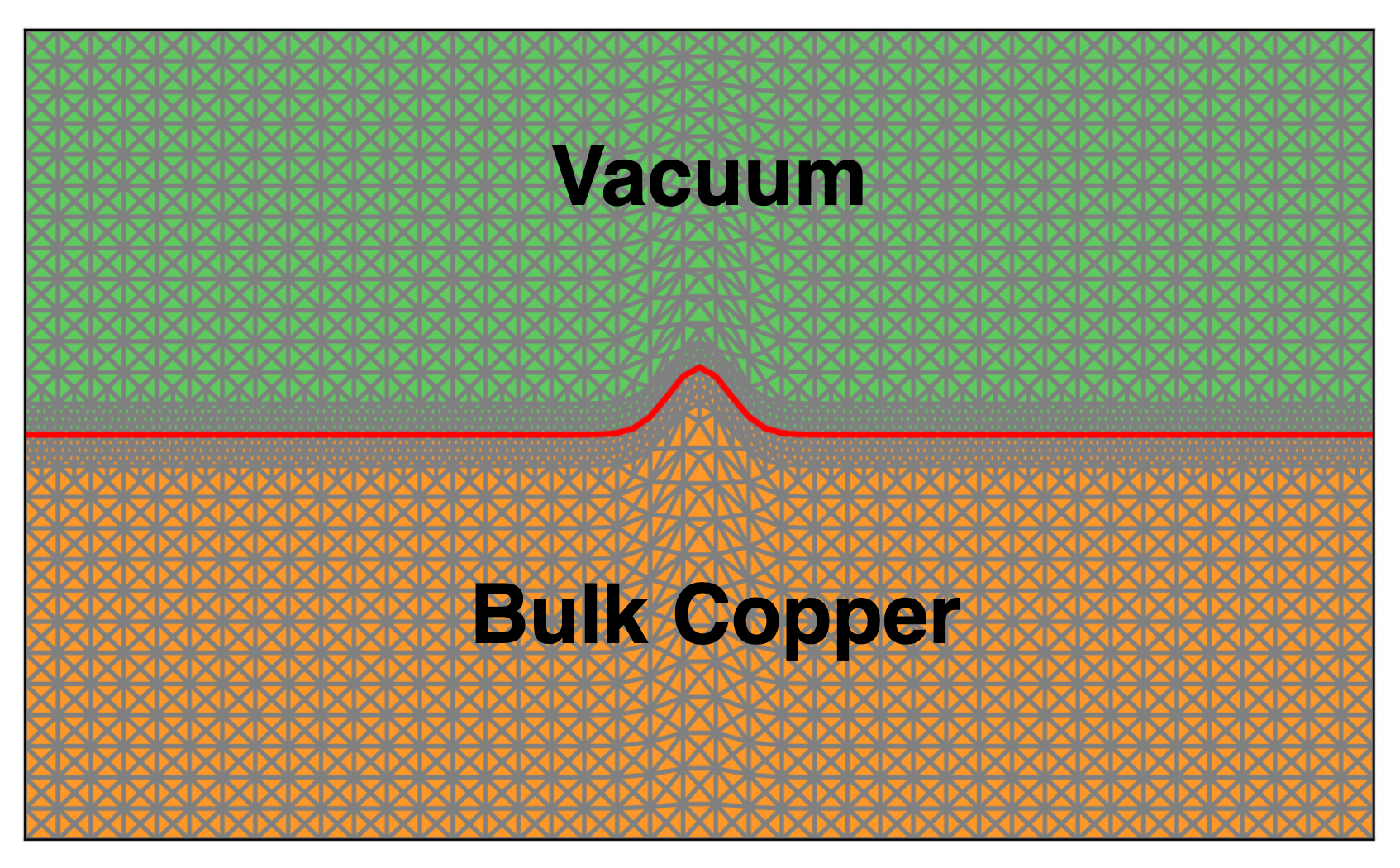}
	\caption{Example of FEM mesh used for the simulation where the vacuum medium is shown as the green mesh and the bulk copper material is shown here as orange. The red line shows the vacuum-copper boundary}
	\label{fig1}
\end{figure}
Consider 2D domains that contain a vacuum and an elastic half-space that here corresponds to bulk copper (Fig.~\ref{fig1}). In this domain, surface evolution  follows from surface diffusion driven by electrostatic, thermo-elastic, and surface energy contributions. In the following, we provide details of the multi-physics workflow which accounts for diffusive evolution of surfaces under combined effects of temperature-increase and electric fields. 

\subsection{Electrostatics analysis}
In order to determine the local electric-field ($E$) along the materials surface, we solve for the electrostatic potential ($\phi$) using Gauss's law in the domain shown in Fig.~\ref{fig1} under an applied electric field ($E_{a}$):
\begin{equation} \label{eq:1}
\begin{aligned}
    \nabla \cdot (\nabla \phi) = 0\\
\end{aligned}
\end{equation}

The material (copper) is idealized as a perfect conductor where the electric potential vanishes, enforcing Dirichlet boundary condition at the copper/vacuum interface. At a large enough distance away from the material surface, we employ Neumann boundary conditions where the field is constrained to the nominal applied field $E_{a}$. Once Eq.~\ref{eq:1} is solved, the electric field at any position can be obtained by taking the negative gradient of the potential $- \nabla \phi = E$.
Although the motivation of our model is to capture surface evolution in the context of rf breakdown, the applied field is kept constant during the simulation, since the time scale for surface morphology evolution is much slower than that of an rf cycle (of roughly 100 ps in X-band.) In this context, we can assume effective dc-field conditions at the surface corresponding to the RMS of an rf cycle.

\subsection{Thermo-elasticity analysis}
To capture the effects of rf losses, we model the elastic deformation in the copper domain resulting from a uniform temperature rise. Such uniform temperature approximation stems from the fact that the high thermal conductivity of copper quickly equilibrates temperature gradients over the micron lengthscales simulated here. Indeed, experiments with rf pulse heating show that temperature gradients tend to form on the scale of $\si{\milli\meter}\sim\si{\centi\meter}$\cite{2}, which is consistent with our assumption. 

The governing equations for small deformations on isotropic and homogeneous materials can be written as in Refs. \cite{18,19}:
\begin{subequations}
\begin{align} \label{eq:2a}
    - \nabla \cdot \sigma(u) = 0 \\ \label{eq:2b}
    \sigma(u) = \lambda tr(\epsilon(u))\mathbb{1}+2\mu \epsilon(u)-\frac{\alpha Y}{1 - 2\nu}\Delta T \mathbb{1}\\ \label{eq:2c}
    \epsilon(u)=\frac{1}{2}(\nabla u + (\nabla u)^{T})
\end{align}
\end{subequations}
where $\sigma$ is the stress tensor, $u$ is the displacement vector, $\epsilon$ is the infinitesimal strain tensor, $\lambda$ and $\mu$ are the Lamé’s elasticity parameters, $\alpha$ is the thermal expansion coefficient, $Y$ is the Young's modulus, $\nu$ is Poisson's ratio, and $\Delta T$ is the temperature change from the reference room temperature of 300 K.

We use fixed displacement boundary conditions at the bottom of the  domain and periodic boundary conditions along the lateral boundaries.
The solution of Eq.~\ref{eq:2a} produces the displacement vector field $u$, describing deformation of the material under a given $\Delta T$. With the displacement vector, the corresponding strain and stress tensor can be obtained from Eq.~\ref{eq:2b} and Eq.~\ref{eq:2c}, respectively.

It is important to note that thermo-elasticity only acts as a driving force if the system is mechanically constrained in some directions (perpendicular to the surface in this case). In an accelerator setting where heating from RF losses mainly affects the first few microns 
from the surface, confinement is provided by the rest of the material which remains colder and hence at a shorter elastic constant. Thermo-elastic driving forces would however vanish for uniformly heated systems that are allowed to freely thermally-expand in all directions.

\subsection{Surface Diffusion}

We incorporate material transport through a surface diffusion process,
where the diffusion flux ($J$) is proportional to the surface gradient of the chemical potential ($\mu_c$) as:
\begin{equation}
    J=-\frac{D_s}{k_b T}\frac{\partial \mu_c}{\partial s} \label{eq:3a}
\end{equation}
where $D_s$ is surface diffusivity, $k_b$ is the Boltzmann constant, $T$ is the absolute temperature, and derivative with respect to $s$ is evaluated along the surface.

From mass conservation, the normal velocity $V_n$ of the surface can be expressed using the surface divergence of $-J$. $V_n$ can then be reformulated in terms of $ \partial h(x,t)/\partial t$, the vertical velocity of the surface profile. The latter can be written as \cite{4,21}:
\begin{subequations} 
\begin{align}
    V_n = \frac{D_s \Omega v_s}{k_b T}\frac{\partial^2 \mu_c}{\partial s^2}\label{eq:3b}\\
    \frac{\partial h}{\partial t} = \frac{D_s v_s \Omega}{k_B T}\frac{\partial}{\partial x}[(1+h'_x)^{-1/2}\frac{\partial}{\partial x}(\mu_c)] \label{eq:3c}
\end{align}
\label{eq::evolution}
\end{subequations}

\noindent Here $h'_x$ indicates the derivative with respect to $x$, and $v_s$ is the number of atoms per unit area.

Aggregating the contribution of electrostatics, thermo-elastic stresses, and surface tension, the chemical potential on the surface can be written as \cite{4}:
\begin{equation}
\begin{aligned} \label{eq:5}
    \mu_c = \Omega (\gamma \kappa + \omega_T - U_E) \\ 
    \kappa = -\frac{h''_x}{(1+h_x^{\prime2})^{3/2}} \\
    \omega_T = \frac{1}{2}( \sum_{j=1}^{2}\sum_{i=1}^{2}\sigma_{ij} \epsilon_{ij} )\\
    U_E = \frac{1}{2}\epsilon_0E_{local}^2
\end{aligned}
\end{equation}
Here, $\omega_T$ is the strain-energy density function for linear isotropic materials, $U_E$ is the electric field-energy density on the material's surface, $\gamma$ is the isotropic surface free energy, and $\kappa$ is the local surface curvature. The surface field-energy density and the strain energy density are calculated simultaneously on their respective mesh, and Eq.~\ref{eq:3c} is solved to calculate the vertical velocity of the surface. The meshes for the vacuum and material domains are then updated to conform to the new interface.
The field and thermal stress is recalculated again for each domain after approximately 100 time steps,  when the surface has evolved a sufficient amount. This frequency is adjusted depending on the geometry and the surface velocity. The open source PDE solver FEniCS \cite{fenics} was used to solve the governing equations using the finite-element method. Variational forms are specified in the FEniCS workflow in the high-level Python-based Unified Form Language (UFL) \cite{varcompiler}. These forms are then automatically compiled and executed through high-performance computational kernels using the finite element library DOLFIN \cite{dolfin}. This computational approach was incorporated in a framework called \textit{SurFE-XD} (\textit{Surface curvature-driven Finite Elements model for Diffusion under eXtreme conditions}, previously introduced to investigate electric-field-driven surface evolution  \cite{electrodiffusion2023}).

\subsection{Initial surface morphology}
We consider three different types of surface geometries: isolated perturbations, sinusoidal perturbations, and random surfaces. An example of isolated perturbation is shown in Fig.~\ref{fig1} where the simulated initial profile is modeled by a Gaussian on an otherwise flat surface. For sinusoidal surfaces, the material's surface is perturbed by a perfect sine wave with constant wavelength $\lambda$ commensurate with the size of the domain. For both Gaussian and sinusoidal surfaces, the aspect ratio of the perturbation is defined as the ratio of the height over either two lateral standard deviation (for isolated perturbations) or the wavelength (for sinusoidal perturbations) of the respective initial surface profile.

Finally, random surfaces were used to capture the evolution of realistic copper surfaces. To model random rough surfaces, the vacuum-copper boundary was represented as a sum of Fourier modes \cite{3}:
\begin{equation} \label{eq:6}
    h(x)=\sum_{n}^{N}A_n sin(n\omega x + \phi_n)
\end{equation}
where the amplitude $A_n$ is obtained by fitting an exponential decay to an experimentally measured spatial frequency spectrum of electro-polished copper photocathodes \cite{1}, and the phases $\phi_n$ are sampled randomly.

\subsection{Linear Stability Analysis}

Linear stability analysis can be utilized to complement fully nonlinear numerical solutions by providing analytical solutions in the limit of small perturbations. Hence, it serves as a baseline verification of the numerical results.

By inserting Eq.~\ref{eq:5} into Eq.~\ref{eq:3c} and assuming a small initial perturbation $h(x) = h_0 \sin(kx)$, the solution  in the frequency domain takes the form of $h(k,t)$=$h_0 \exp(g(k)t)$ where $g(k)$ is the growth rate corresponding to wavenumber $k$. In the small perturbation regime, Eq.~\ref{eq:5} can be simplified as $\kappa\approx\frac{\partial^2 h}{\partial x^2}=-Ak^2sin(kx)$, $U_E$=$\epsilon_0 E_0^2(ksin(kx))$, $\omega_T=2kY\alpha^2 \Delta T^2$, resulting in the temporal surface evolution described as \cite{22,23}:
\begin{subequations}
\begin{align}
    \frac{\partial h}{\partial t}=C[-\gamma h''''_x-k h''_x(2\alpha^2 \Delta T_0^2 Y-\epsilon_0 E_0^2)] \label{eq:7a} \\
    h(t,k)=h_0 \exp[C(2k^3\alpha^2\Delta T_0^2 Y + \epsilon_0 E_0^2 k^3 - \gamma k^4)t] \label{eq:7b}
\end{align}
\end{subequations}
where $C=\frac{D_s v_s \Omega^2}{k_BT}$. 

\begin{figure}
	\includegraphics[width=8.6cm]{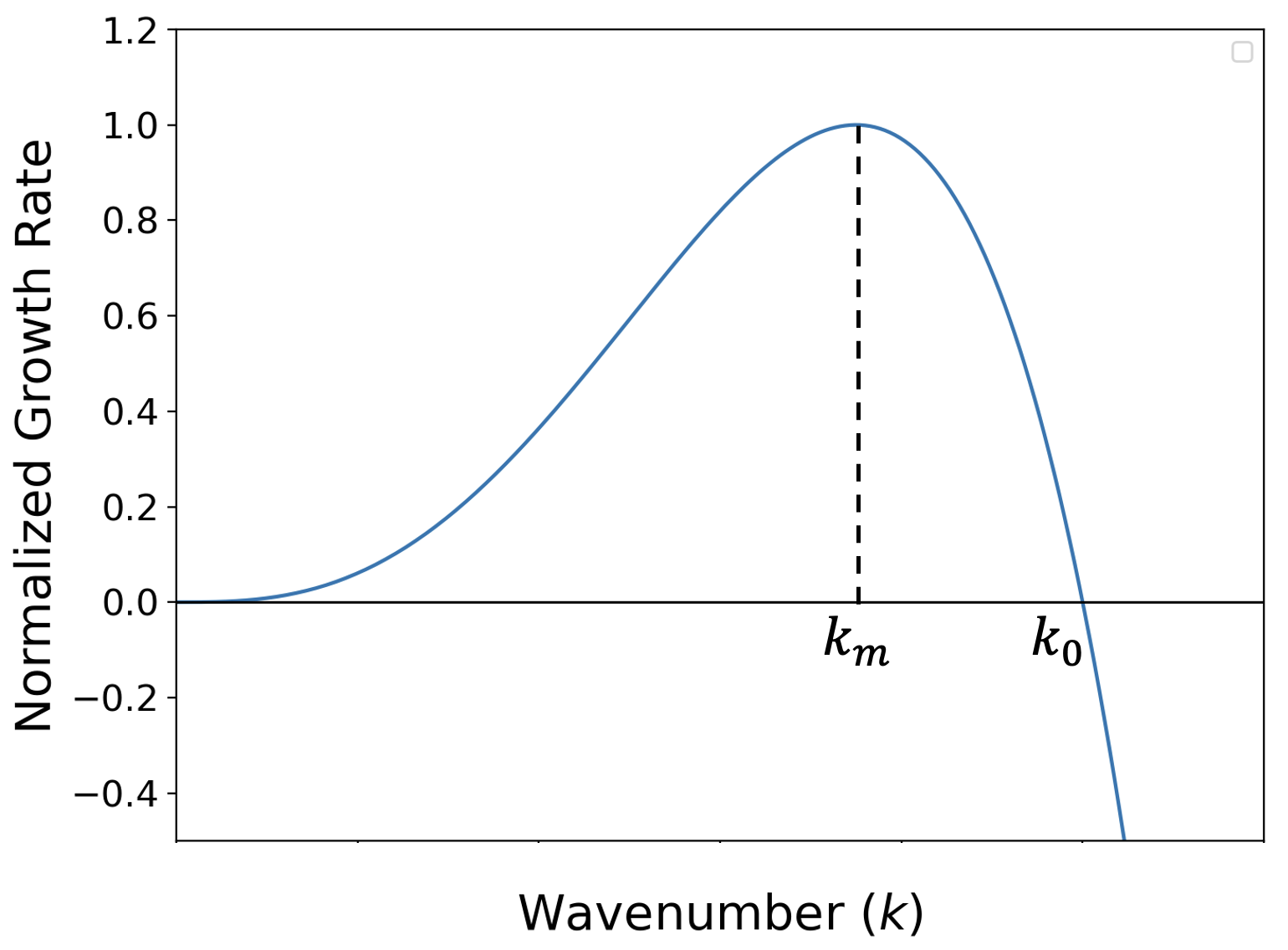}
	\caption{Growth rate $g(k)$ for surface wavenumber $k$. Perturbations with wavenumber $k<k_0$ will grow while those corresponding to larger wavenumbers will decay. The maximum growth rate occurs at $k=k_m$. See text for details.}
	\label{fig_growth}
\end{figure}

Fig.~\ref{fig_growth} reports the growth rate $g(k)=C(2k^3\alpha^2\Delta T_0^2 Y + \epsilon_0 E_0^2 k^3 - \gamma k^4)$. The result shows that perturbations corresponding to wavenumber below a critical value $k_0=1/\gamma (2\alpha^2\Delta T_0^2 Y +\epsilon_0 E_0^2)$ will grow, while larger wavenumber modes will decay. The fastest growing mode can be found by solving $\partial g(k)/\partial k=0$, which gives $k_m=3 k_0/4$. The result indicates that the maximally unstable mode $k_m$ and critical mode $k_0$ both depend quadratically on the operating/applied field and temperature rises. In general, higher temperatures or applied fields allow for the spontaneous growth of perturbations at increasingly large wavenumbers (increasingly small wavelengths). For typical applied field of 100 MV/m and $\Delta T$=100 K, the critical wavelength, $\lambda_0=2\pi/k_0$, is approximately 12~$\mu$m. If the gradient increases to 300 MV/m, $\lambda_0$ decreases to  6~$\mu$m.

Eq.~\ref{eq:7b} can be used to approximate small amplitude surface evolution by taking the Fourier transform of the equation.  The critical temperature rise, $T_{cr}$, and the critical nominal field, $E_{cr}$, can be defined using this approximation by finding the value in which linear stability approximation predicts a zero growth velocity. It is important to note that this approximation is only expected to hold for small perturbations, as interactions between modes are not taken into account by this approach.

\section{Results}

\subsection{Effect of heating}
\label{subsec::heat}

\begin{figure}
	\includegraphics[width=8.6cm]{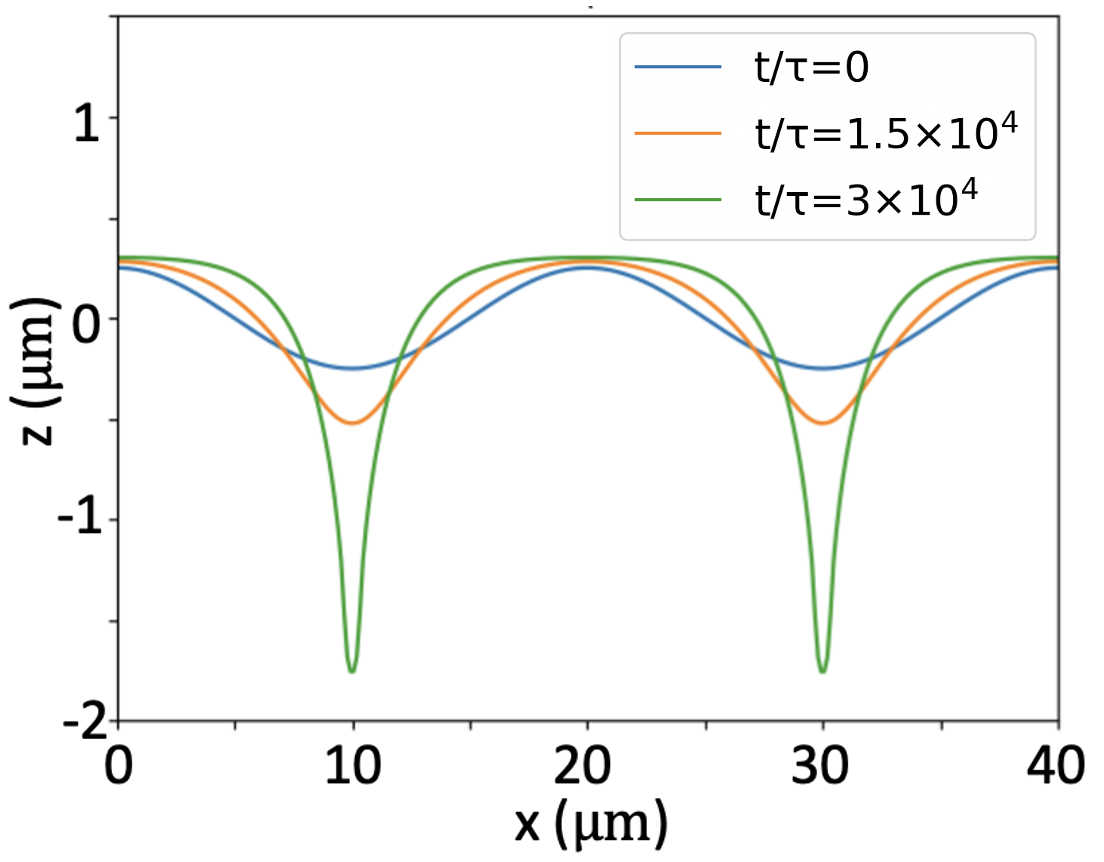}
	\caption{Simulated copper surface profile after applying a $\SI{120}{\kelvin}$ uniform temperature rise to an initial sinusoidal perturbation.} 
	\label{tempfig}
\end{figure}

In the following, the temporal evolution of the modeled surface profiles are reported in dimensionless time $\tau$. Conversion to physical time units will be discussed in the Section~\ref{Diss}.

We first consider the evolution of a copper surface subjected to a uniform temperature rise in the absence of applied electric fields \cite{4}. 
The results of a uniform temperature rise of 
120 K applied to a sinusoidal surface is shown in Fig.~\ref{tempfig}. A temperature rise in the bulk material results in compressive stresses due to the material being constrained in the lateral direction. The resulting deformation causes the strain-energy density to be concentrated at the trough of the sinusoidal profile, resulting in sharpening of the troughs and simultaneous broadening of the peaks. These concomitant sharpening and broadening occur due to conservation of mass during the surface diffusion: the material removed from the trough is redistributed to the surroundings of the original maximum. This is a key point -- the evolution of a small amplitude perturbation dominated by temperature rise alone is unlikely to drive the formation of a type of precursor that would lead to breakdown, as flat peaks and sharp troughs do not couple efficiently with an electric field in a way that is expected to cause runaway field emission. Electric fields are therefore expected to be a critical ingredient in the formation of BD precursors.


\subsection{Effect of electric field}
\label{subsec::field}
In contrast, the application of sufficiently large electric fields has been shown to lead to runaway tip growth and sharpening, due to the localized enhancement of electric fields in regions of high negative curvatures \cite{electrodiffusion2023}. This mechanism was observed to lead to the formation of breakdown precursors on copper surfaces at fields on the order of 250 to 500 MV/m, depending on the characteristics of the initial surface perturbations.
Grooving instabilities were not observed under electric field alone. 
Similar mechanisms drive the formation of Taylor cones in liquid conductors \cite{TaylorLiquid} as well as on metal electrode surfaces \cite{Srolovitzelectrostatic,Experiment_surface_electric}. Therefore, field driven evolution of initially small-amplitude perturbations initially 
present on the metal surface is a likely candidate for the formation of sharp asperity breakdown precursors.

\subsection{Critical $E$-field and heating regimes for instability}
\begin{figure}
	\includegraphics[width=8.6cm]{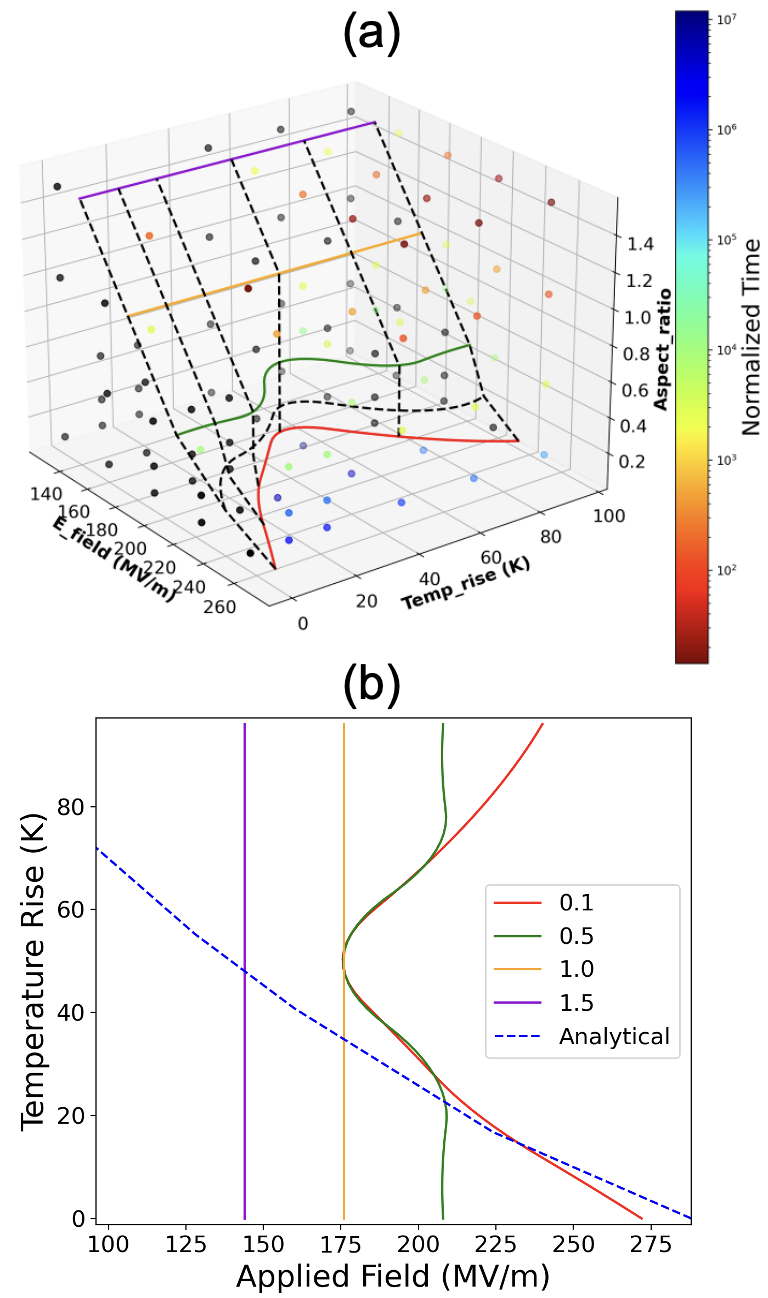}
	\caption{(a) Phase Diagram showing the normalized time when initial surface profile grow a 10$\%$ of the original height or reach instability for given combination of parameters. (b) Interpolated horizontal slice of the phase diagram at various aspect ratio with analytically predicted boundary for low aspect ratio.} 
	\label{fig2}
\end{figure}

In order to unveil the synergistic effect between the two driving forces, i.e. a combined thermal and electric field effects, we conducted series of simulations by applying electric field ranging from 140-275 MV/m 
and temperature increase ranging from 0-100 K 
to act on a surface profile having an aspect ratio ranging from 0.1-1.5. Here, the initial asperity has a Gaussian profile with a 2 standard deviation width of 10 $\mu$m. 


Fig. \ref{fig2}(a) reports a phase diagram of simulations with different combinations of the aspect ratio, applied field and temperature augmentation. Here, the color scale represents the time at which the surface grows by 10\% of the initial height or reaches instability at the peak $k_m$, i.e. moves in the supercritical regime (Fig.~\ref{fig_growth}). The instability is characterized by the acceleration of the surface velocity at the peak of profile when the surface reaches instability or when elapsed computational time $t$ is 1 s in SI units. The black points represents the simulations in which called operating temperature and electric field were not sufficient to reach instability runaway, i.e. resulting in the initial protrusion to decay back towards flat surface. The black dashed lines and the colored lines in Fig \ref{fig2}(a) separate these two regimes, but should not be taken as the precise location of the actual boundary, given the limited number of simulations.

\begin{figure*}
	\includegraphics[width=\textwidth]{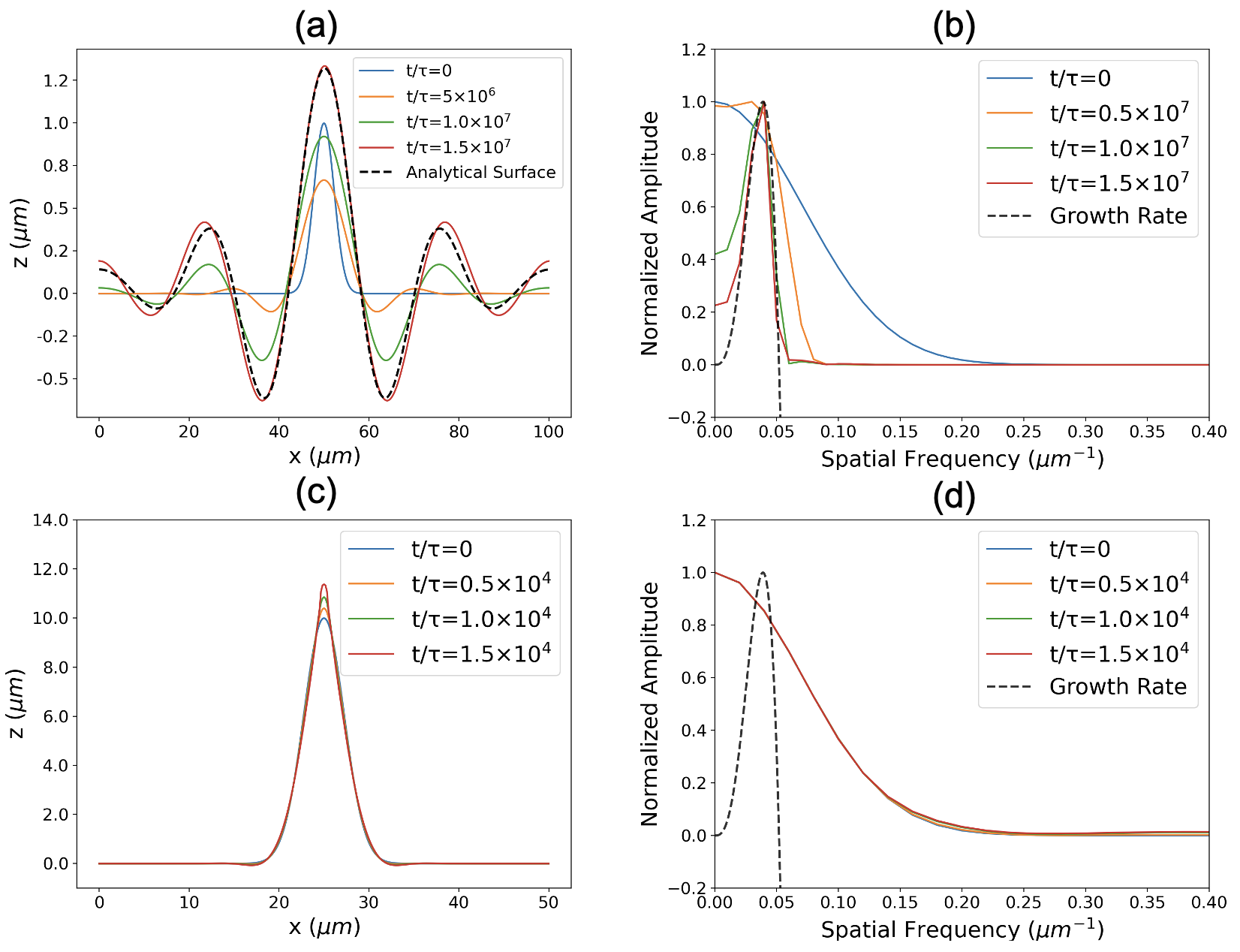}
	\caption{(a) Surface evolution with $\SI{48}{\kelvin}$ temperature rise, 192 MV/m, and 0.1 aspect ratio. (c) Surface evolution with $\SI{48}{\kelvin}$, 192 MV/m, and 1.0 aspect ratio.
	(b,d)  Fourier transform of the surface profile from the FEM simulations (b) with analytical growth rate curve calculated from Eq. \ref{eq:7b}}
	\label{fig3}
\end{figure*}

At small aspect ratios (0.1 (red) and, to a lesser extent 0.5 (green)), temperature and applied electric field display synergistic effects, causing tip growth in a regime where electric fields alone would be sub-critical. This behavior is initially captured by the linear stability analysis introduced in the Section~\ref{Fab}.E above (blue dotted line), within the uncertainty of the exact location of the boundary. As the temperature rises above a certain value (about $\SI{55}{\kelvin}$), the trend reverses and heating contributes less and less to the formation of breakdown precursors. In contrast, at higher aspect ratios ($\ge 1$), the propensity for BD precursor formation is unaffected by heating (as indicated by the vertical lines). 



Note that high amounts of heating alone do not lead to BD, as it instead leads to grooving instabilities with blunt peaks, as described above. The resulting reduction in peak (negative) curvature creates microstructures that couple weakly with the electric field, and hence to the absence of tip growth. Therefore, breakdown would not be expected in the high-$T$/low-$E$ regime, as observed in Fig.~\ref{fig2}.

The opposite regime, high-$E$/low-$T$ has been previously investigated \cite{electrodiffusion2023}. In this case, the critical electric field required for instability was shown to decrease with aspect ratio, i.e., a consequence of the fact that the electric field enhancement factor at the tip is linearly proportional to the aspect ratio. This effect is observed in Fig.~\ref{fig2}, as expected.

On general grounds, one might also expect that heating would not affect tip growth at high aspect ratio, as the elastic energy due to thermal loading 
can be efficiently relaxed since the material is not laterally constrained at the tip (in contrast to bulk materials or low aspect ratio asperities). In this limit,  BD propensity can be expected to be unaffected by heating. This is again consistent with the results shown in Fig.~\ref{fig2}.


The intermediate regime of $E$ and $T$ at low aspect ratio is more complex.
Fig.~\ref{fig3}(a) illustrates a surface profile for a simulation with initial aspect ratio of 0.1 under application of a field of 192 MV/m and a temperature augmentation of 48~K. This corresponds to conditions where BD would not have been expected
under this electric field alone; heating is therefore enhancing BD precursor formation in this case. 
The analytical solution of Eq.~\ref{eq:7b} under this small surface perturbations is plotted against the simulation results. The agreement is excellent, thereby showing that the emerging wavelength, shown in Fig.~\ref{fig3}(b), follows the predictions of linear stability analysis. Additionally, the spectral analysis in Fig.~\ref{fig3}(b) illustrates the strong selection of a particular surface wavelength corresponding to the maximum of the predicted growth rate $g(k)$ for the applied parameters combination. This amplification process is efficient as the initial surface profile has a significant amplitude at that frequency. Higher frequency modes are quickly suppressed, leading to an efficient wavelength selection process.
The mode amplitude obtained computationally are in excellent agreement with the analytically predicted growth rate. Analytically, the fastest growing mode $k_m/(2\pi)$ is predicted for a wavelength of 0.039~$\mu$m$^{-1}$, compared to 0.04~$\mu$m$^{-1}$ observed in the simulations. When the aspect ratio is much higher, (1.0, which is a case that would be expected to experience BD under the corresponding electric field alone) Fig.~\ref{fig3}(c,d) show that wave-like surface evolution with a strong wavelength selection does not occur (at least initially). Instead, high frequency modes are amplified by coupling with the $E$-field at the tip, leading to rapid tip growth and sharpening, and ultimately to BD. As shown by the comparison to the analytically predicted growth rates (dashed line), the evolution at high aspect ratio cannot be described by linear perturbation and is instead dominated by non-linear effects caused by the large curvature at the tip, 
which lead to the growth of high-frequency modes.


\begin{figure*}
	\includegraphics[width=\textwidth]{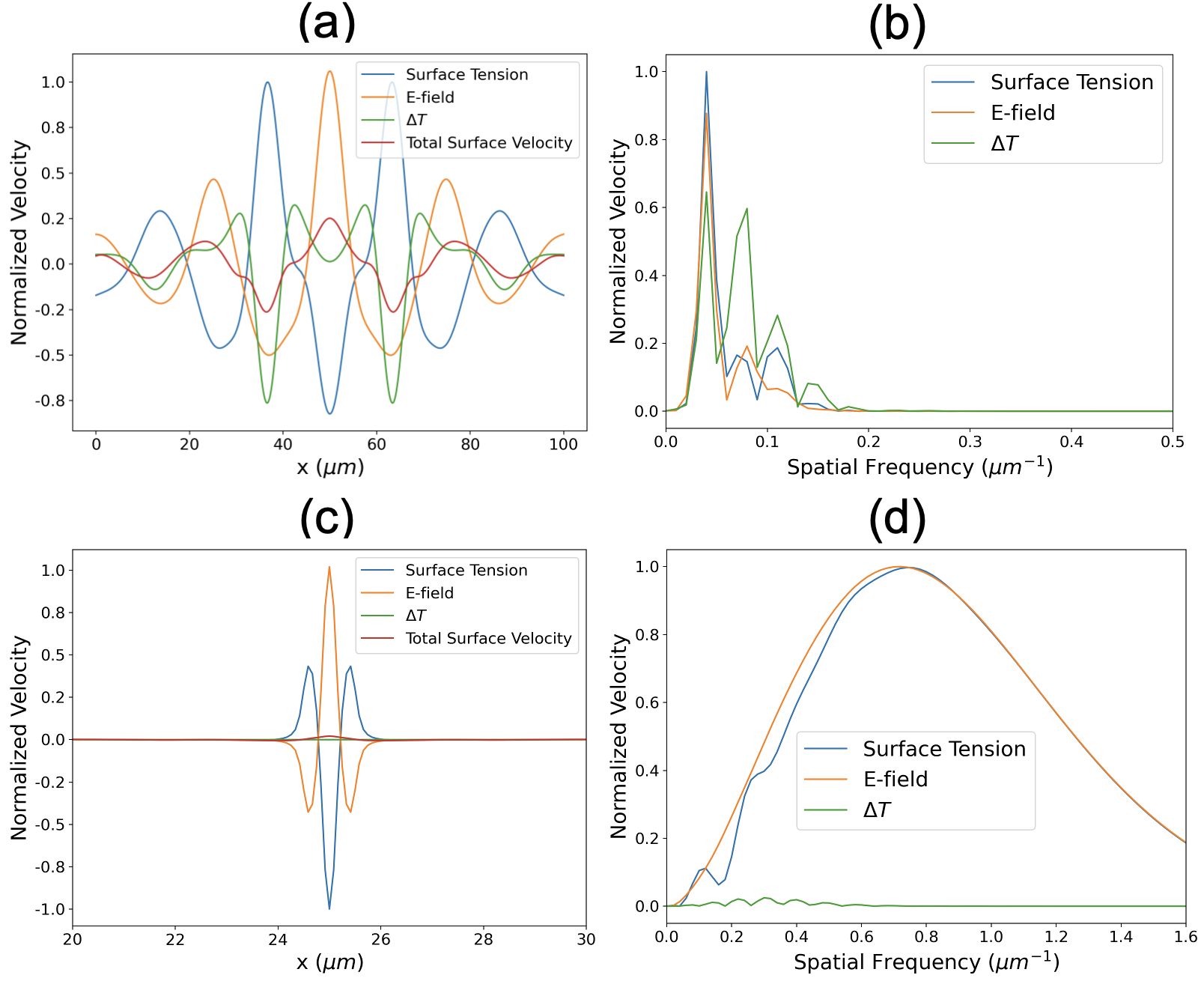}
	\caption{(a) Velocity components of the final surface profile shown in FIG. \ref{fig3}(a) for surface tension, applied electric field, and temperature rise. The applied parameter for the simulations are 0.1 aspect ratio, $\SI{48}{\kelvin}$ temperature rise, and 192 MV/m. (c) Velocity component for simulation with 1.0 aspect ratio., $\SI{48}{\kelvin} \Delta T$, 192Mv/m. (b,d) Fourier Transform of the individual velocity components respectively for the final surface in (a) and (b).} 
	\label{fig4}
\end{figure*}

In order to elucidate the mechanics behind the surface evolution in a coupled setting, it is useful to isolate
 the surface velocity contributions stemming from
surface tension, electric field, and heating. Such a summary is shown in Fig.~\ref{fig4}(a,d) for two aspect ratios, 0.1 and 1.0.


The surface tension contribution (blue line) is observed to counteract any curvature: when negative curvature develops (at tips), the surface tension's contribution is negative, leading to blunting; in contrast, when the curvature is positive (at grooves), the surface tension component is positive, again leading to blunting. In contrast, the electric field contributions (orange line), tend to amplify both tips and grooves, as the material diffuses along the field gradient (from grooves where the applied field is partially shielded towards tips where the field is amplified). Finally, heating (green lines) tends to amplify grooves much more than tips, as discussed below. At low aspect ratios, the velocity contributions from each driving forces are similar. In contrast, at higher aspect ratios, corresponding to Fig.~\ref{fig4}(d), the relative contribution from heating is marginal compared to the  surface and electric field components, consistent with the above discussion that thermo-elasticity couples weakly with tip-like structures, in contrast to electric fields that are strongly amplified at such features \cite{electrodiffusion2023}. In this case, the surface evolution is only weakly affected by heating, consistent with the results shown in Fig.~\ref{fig2}.

\begin{figure}
	\includegraphics[width=8.6cm]{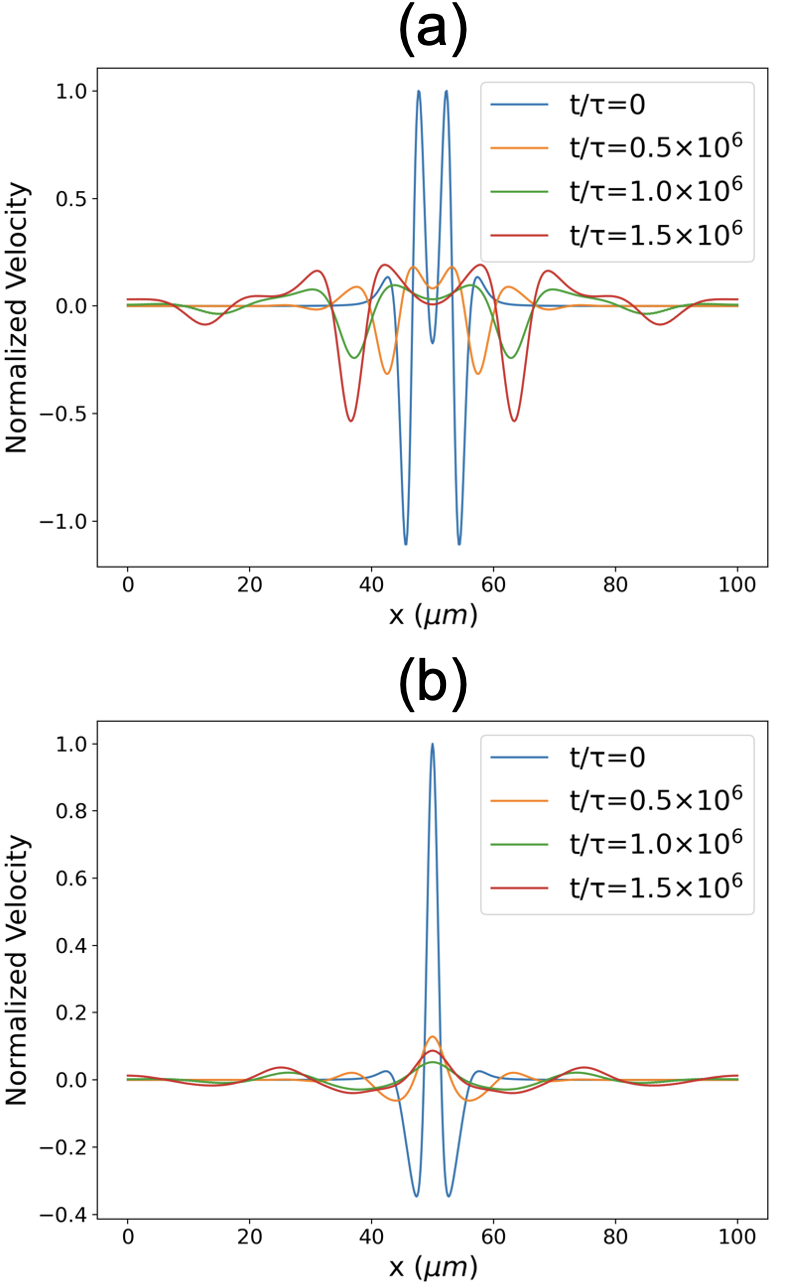}
	\caption{Velocity contribution from (a) temperature rise and (b) applied field at various $t/\tau$ for the surfaces shown in Fig. \ref{fig3}(a)}
	\label{TempFieldvelocity}
\end{figure}

Electric fields therefore provide the main driving force leading to the increase in negative curvature necessary for BD to occur. Nonetheless, Fig.~\ref{fig2} clearly shows that thermo-elastic driving forces can lead assist tip growth when acting concurrently with an electric field. At low heating, the observed enhancement follows the predictions of the analytic linear perturbation model. This suggests that the main effect in this case is the amplification of the growth rate of high frequency modes that are required for the eventual creation of sharp features that efficiently couple with the electric field to cause BD. This is confirmed in Fig. \ref{TempFieldvelocity}(a), where it can be seen that thermo-elasticity  contribute a relatively small but positive velocity at the tip location at short times. As the aspect ratio of the asperity increases, this velocity component correspondingly decreases as expected.

As the temperature increases further, the trend reverses and the assistance provided by heating becomes less significant. Departure from the linear-stability predictions suggest that non-linear/large-deformation effects are responsible for this reversal. The behavior observed for strong heating in Fig. \ref{tempfig} shows that runaway groove growth (which form in the non-linear, large amplitude, regime) leads to a blunting of the tip region. The fact that the location of the initial tip moved up during surface evolution shows that blunting resulted from the mass flowing out of the groove, and not primarily from surface tension (which would have pulled the tip region down). The same effect appears to be responsible for the decrease in assistance provided by a large amounts of heating with respect to BD formation at small aspect ratio. Our simulations show that if runaway groove growth sets in before tip growth, the mass flowing out of the grooves contributes to a decrease in the curvature in the tip region that hampers the sharpening of the tip, hence leading to a relative decrease in the efficiency of the coupling with the electric field, and a relative decrease of BD propensity. Note that even beyond the turnover point, thermo-elasticity still contributes to a decrease in the critical electric field for BD, although this enhancement is lessened for very large temperature rises.

\subsection{Random Surface Simulation}
\begin{figure*}
	\includegraphics[width=\textwidth]{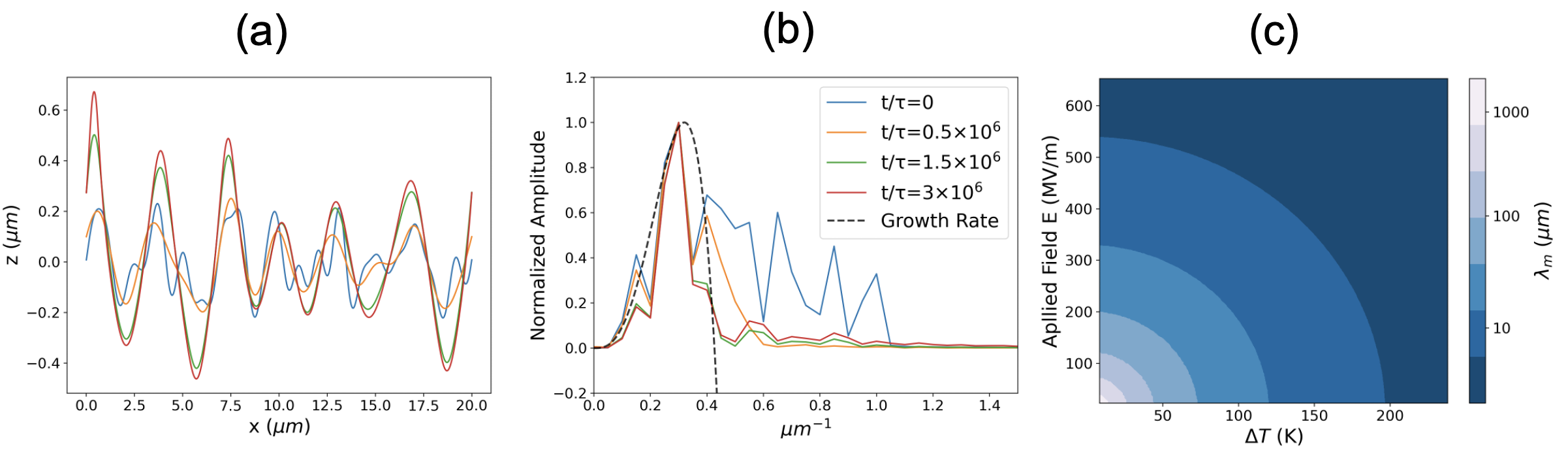}
	\caption{(a) Surface evolution of the random surface under application of electric and thermal gradients of 460 MV/m and 130 K. The initial surface structure was initialized following the frequency content observed experimentally for polished copper surfaces \cite{1}.
 (b) Fourier transforms of the surface profiles as compared to the growth rate $g(k)$. (c) Log scale contour plot showing the fastest growing wavelength $\lambda_m$ for given combination of applied field and $\Delta T$.}
	\label{fig5}
\end{figure*}

The analysis above focused on isolated perturbations. In realistic surfaces, roughness will initially be present on a wide range of scales Fig.~\ref{fig5}. In the following, realistic surface profiles were generated using the amplitude/wavelength characteristics measured on actual copper photocathode surfaces \cite{1}.

Field gradient of 460 MV/m and $\Delta K$ of 130K were applied on the copper surface to simulate its morphology.
The simulation result and growth behavior of the surface remains consistent with the linear stability analysis and the previous result for isolated geometry.
The evolution of the frequency spectra in the simulations show the preferential growth of the maximally unstable frequency modes, while high frequency modes above the critical frequency $k/2\pi>k_0/2\pi$, sharply decay. The combination of gradient and $\Delta T$ caused the surface to morph into a new more regular wave-like surface as presented by Fig. \ref{fig5}(a,b). 

This wavelength selection is in excellent agreement with the prediction of the linear stability analysis.
The peak in the spatial frequency spectrum forms at $0.3~\mu \text{m}^{-1}$ which is very close to the maximal unstable frequency $1/\lambda_m$ = $0.32~\mu \text{m}^{-1}$ as calculated analytically from Eq.~\ref{eq:7b}.

As shown, the linear stability analysis gives an excellent prediction for the growth behaviour of copper surfaces in the small amplitude (aspect ratio) regime. To understand the relation between the maximally unstable mode $k_m$ and the applied parameters, Fig.~\ref{fig5}(c) charts the evolution of maximally-unstable wavelength $\lambda_m$ against given combination of applied gradient and $\Delta T$. The results show that $\lambda_m$ rapidly decreases from 1,000 to 10 $\mu$m as temperature increases by 200 K and the gradient increases to 300 MV/m. The decrease in $\lambda_m$ becomes more gradual at larger applied parameters. For instance, $\lambda_m$=$2~\mu m$ under 600 MV/m and 230 K condition. This analytical treatment provides useful guidelines for the analysis of surfaces exposed to combinations of electrostatic and thermo-elastic driving forces, as a robust wavelength selection would constitute the strongest experimental signature of the effects discussed here. 


\section{Discussion}\label{Diss}

The simulation results presented above suggested that surface diffusion can be an effective mechanism that contributes to breakdown precursor formation.
To predict the physical timescale over which this mechanism can be expected to occur, the simulation time unit ($\tau$) are related to physical times as follows:
\begin{equation}\label{eq:8}
    \tau = \frac{k_b T}{\gamma D_s v_s \Omega^2}.
\end{equation}
The main unknown in this conversion factor is the effective diffusivity of copper atoms on surfaces,
which, on clean surfaces is of the form $D_s=3.615^2/2\times10^{12}\exp(-E_b/k_bT) \: \AA^2/$s \cite{butrymowicz1973diffusion},
where $E_b$ is the migration barriers, and a standard prefactor of $10^{12}$/s was assumed. According to both experimental and theoretical calculation, 
the migration barriers vary considerably for different surface orientations: 0.1-0.15 eV for (111), 
0.25-0.30 eV for (110), and 0.38-0.69 eV for (100). \cite{24,25,26,27,28,29} 
Using $\gamma=1.5$ J/m$^2$ \cite{kumykov2017surface}, $\Omega=11.81\:\AA^{3}$, $v_s=0.306 \: \AA^{-2}$ \cite{simon1992properties}, this wide range of possible diffusivities translates into 
a broad range of physical timescales $\tau$ varying from 4.75 ms ($E_b= 0.1$ eV) to 11 seconds ($E_b=0.3$ eV) with room temperature. 
Due to the exponential dependence of the diffusivity with temperature, a temperature rise can significantly decrease the timescale, e.g., with $\Delta T$ of 100 K, the characteristic timescale vary from 2.4 ms ($E_b= 0.1$ eV) to 0.81 seconds ($E_b=0.3$ eV). Further, the growth rates of perturbations can vary by orders of magnitude depending on the applied fields and temperature rises, as shown in Fig.\ \ref{fig2}. From the linear stability analysis described above, one can indeed show that the growth rate of the maximally unstable mode will increase as $\mathcal{O}[\exp (-E_b/k_b(T+\Delta T_0))(E_0^2 + \Delta T_0^2)^4]$. This is qualitatively consistent with the dramatic acceleration of the breakdown rate with increasing gradient and temperature observed in the literature \cite{simakov2018advances} (where it was observed to scale with applied field to a large power \cite{descoeudres2009dc} ). Non-linearities due to high-aspect ratios are expected to further increase the breakdown rates \cite{electrodiffusion2023}. 



Our results are consistent with a number of experimental observations. First, the peak pulse heating temperature is known to be strongly correlated with the breakdown rate \cite{simakov2018advances}. This is consistent with the synergistic effect of electric fields and heating in increasing the breakdown rate, and with the exponential increase of the surface diffusivity with temperature, which is also expected to reduce to time required for precursors to form, as discussed below. Second,
the localization of breakdown events (i.e., the propensity of breakdown craters to overlap with one another) observed in pulsed DC experiments \cite{PROFATILOVA2020163079} strongly suggests that debris and craters created by previous BD events are efficient nucleation sites from subsequent BD events. This is consistent with the results shown in Fig.\ \ref{fig2}, as preexisting  large amplitude roughness leads to the formation of further BD precursors at low fields and in short amounts of time. Further, it could be expected that breakdown precursors would develop faster on Cu(111) facets of polycrystalline surfaces. Indeed, it was found by electron back-scattering diffraction measurements of conditioned Cu samples \cite{2} exposed to cyclic pulsed heating (without a significant electric field applied), that the resulting surface roughness was heavily dependent the orientations of the surface. The surface orientation with the largest damage/roughness development was observed to be (111), followed by (110), and finally by (100). This experimentally distinct difference in the extent of roughness and damage evolution is consistent with the predictions of our model which predicts a very strong dependence of the rate at which surface features would grow depending on the surface migration energy. We however note that this specific experiment was carried out under conditions of thermal cycling, so other mechanism (e.g., the formation of shear bands) could have contributed to the development of surface roughening in this case.

\section{Conclusion}\label{Conc}
In this work, we presented a numerical model describing how metallic surfaces undergo tip-growth instabilities through surface diffusion under combined applied electric fields and thermal stresses. Such instabilities can lead to the formation of geometrically sharp features that can act as
breakdown precursor. 
The results show that thermal stressed induced by temperature rises of a few tens of Kelvin (which are typical of conditions caused by Joule heating in copper accelerator cavities) can significantly lower the electric fields needed for breakdown precursors to form compared to situations where heating is absent. Our results show that diffusion-driven surface evolution can spontaneously create breakdown precursors in typically accessible regimes of applied fields and temperature rises for accelerator applications. Surface diffusion is therefore a viable candidate mechanism to explain the formation of sharp breakdown precursors.


\section*{Acknowledgments}
Ryo Shinohara, Soumendu Bagchi, Evgenya Simakov, and Danny Perez were supported by the Laboratory Directed Research and Development program of Los Alamos National Laboratory under project number 20230011DR.
The work by Sergey Baryshev was supported by the U.S. Department of Energy, Office of Science, Office of High Energy Physics under Award No. DE-SC0020429.
Los Alamos National Laboratory is operated by Triad National Security, LLC, for the National Nuclear Security Administration of U.S. Department of Energy (Contract No. 89233218CNA000001).
\bibliography{references}

\appendix
\addcontentsline{toc}{section}{Appendices}
\renewcommand{\thesubsection}{\Alph{subsection}}

\end{document}